\documentclass[twocolumn,showpacs,preprintnumbers,
amsmath,amssymb]{revtex4}
\usepackage{graphicx}

\begin{document}
\title{DILUTE METALS:                                                                                                        SUPERCONDUCTIVITY, CRITICAL CURRENTS, MAGNETIC PROPERTIES}
\author{V. N. Bogomolov}
\affiliation{A. F. Ioffe Physical \& Technical Institute,\\                       Russian Academy of Science,\\                                                     194021 St. Petersburg, Russia} \email{V.Bogomolov@mail.ioffe.ru}                  \date{\today}                                                                     \begin{abstract}                                                                      Properties of oxides are interpreted as a result of existence of the virtual sublattices formed
    by the atomic quantum states. An infinite cluster with the superconductivity of the Bose-Einstein
    condensate kind can be formed in the ground state sublattice at certain oxigen atoms concentration
    in the effectively diluted system of metal atoms (above the percolation threshold). Then the electron
    pairs concentration $n/2$ can be much less than the metal atoms concentration N in the oxide.
    The similar situation takes place in metals with superconductivity of the BCS type.
    Above the percolation threshold the superconductivity  $T_{c}$  may be limited by the  magnetic
    properties of the oxigen $2p^{4}$ quantum state sublattice. Data on the critical current density
    allows us to estimate the electronic density $n/2$ and to obtain an information on the superconductivity
    nature. \end{abstract}
\maketitle
\bigskip
\paragraph*{\textbf{I.}}  A structure of the metal oxides can be conceived as a superposition of four virtual sublattices
formed

by \; (i) \quad - $yN$ metal atoms in the ground state,

\qquad (ii)\quad - $(1 - y)N$ metal ions,

 \qquad (iii) \: - $(1 - y)N$  $\textrm{O}^{2-}$ iones, \\and \quad (iv) \: - $yN$ paramagnetic oxigen atoms \cite{bib1}).\\ These sublattices are related to the quantum states
of the metal and oxigen atoms and to certain physical properties of the oxides. The value $N$ is
an actual concentration of the metal atoms in the oxide, while the value $y$ is a portion of the
ground state in the superposition of the ground and ionized states of the metal atoms. Such
approach allows us to make several conclusions on superconducting, magnetic and ferroelectric
properties of the oxides.
\paragraph*{\textbf{II}.} An actual density $n$ of superconducting electrons can be estimated from the data on the
critical current density $j_{c}$, if we assume that $kT_{c} \sim mv_{c}^{2}/2$ and  $j_{c} \sim env_{c},$ where $v_{c}$ is the
current carriers critical velocity.
 The critical current density $j_{cL}$ takes values (at \textit{T} and \textit{H} about zero) within the range $(10^{5} - 10^{6}) \, \textrm{A/cm}^{2}$ for low-temperature superconductors (LTS) (Nb-Ta with $T_{cL} \sim 10\textrm{K}$ \cite{bib2,bib3}).
At similar conditions, $j_{cH}$ reaches the magnitudes
$(10^{7}-10^{8}) \, \textrm{A/cm}^{2}$ for high-temperature
superconductors (HTS) (for instance,
$\textrm{YBa}_{2}\textrm{Cu}_{3}\textrm{O}_{7-x}$ with $T_{cH}
\sim 90\textrm{K}$ or layers of $\textrm{CaCuO}_{2}$ intermitted
with $\textrm{Ba}_{0.9}\textrm{Nd}_{0.1}\textrm{CuO}_{2+x}$ layers
with $T_{cH} \sim 60\textrm{K}$ \cite{bib4,bib5}).\\  At
$m=2.5m_{e},\: n_{L} =5.7\times10^{18}\textrm{cm}^{-3},\:
n_{H}=1.8\times 10^{20}\textrm{cm}^{-3}.$

It follows from the BCS model for metals with $N \sim 10^{22}\textrm{cm}^{-3}$ that $n_{L}/2 \sim N(kT_{cL}/E_{F}),$
    where $(kT_{cL}/E_{F}) \sim 10^{-4}.$ In the case of HTS oxides it is necessary to have
    a parameter $(kT_{cH}/E_{F}) \sim 10^{-2}$ in the BCS theory that formally is realistic for $E_{F} \sim 1\textrm{eV}.$
    Then the critical current density $j_{cH} \sim (kT_{cH})^{3/2}/E_{F}.$

         However, one can try the model of Bose-Einstein condensing (BEC) hard electronic
    pairs in both of cases as well. Then $kT_{cB} = 3.3\times(h/2\pi)^{2}(n/2)^{2/3}/2m^{*}$ and $kT_{cBL} \sim 12\textrm{K},$
    while $kT_{cBH}  \sim 120\textrm{K}.$ Both of temperatures are near to experimental magnitudes
    (\!$10\textrm{K}$ and $90\textrm{K}$\!) in this version too. In other words, knowledge of only the transition
    temperature magnitude is not suffice to discriminate between the BCS and BEC mechanisms.

       Data on critical currents can give an additional information on the nature of transition
   into the superconducting state. In the BEC case the current $j_{c} \sim (kT_{c})^{2}$ and all electrons
   participate in it unlike the BCS metal case, where their participation ratio is $(kT_{c}/E_{F}).$
   (The condition $j_{c} \sim (kT_{c})^{2}$ corresponds to the relation $kT_{cB} \sim kT_{cF}  \sim E_{F}).$ This fact excludes BEC as a mechanism of the superconducting transition for metals, but can be an argument
for applicability of this model at least in a restricted region taking into consideration that metals
in oxides exist in an diluted state ("chemical" dilution \cite{bib1}). Certainly, a real hundredfold dilution
is not accessible, but an arbitrary effective dilution degree can be reached in the infinite cluster
of metal atoms ground states. New collective properties appear, when an average effective density of
ground states exceeds the percolation threshold $y_{ð}$  for a density of quantum states per metal
atom (Fig.1). The parameter \textit{y} is proportional to the similar parameter \textit{x} in the chemical
formula $\textrm{YBa}_{2}\textrm{Cu}_{3}\textrm{O}_{8-x}.$

    Above the percolation threshold for forming of the superconducting state, the ground
    states associate into the infinite cluster with the electronic pair density $n_{c}/2,$
    where $0<n_{c}<N$ that leads to forming of the superconducting state. The metal atoms
    ground states concentration in the cluster (the order parameter) grows with \textit{y},
    while the effective interatomic distances and effective electron masses decrease (Fig.1).
    Tunneling of the atomic electron pairs can lead to their BEC. However, further increase
    of $n_{c}$ leads to a drop of the electronic pair  binding energy   and, at some
    magnitude of $n_{cc},$ it would be favorable for the cluster electrons to transit
    into the Fermi state, i. e. into the metal or insulator state. The BEC - BCS
    crossover condition  $\Delta \sim 5.5 kT_{cB}$ can be estimated setting the electron energy
    density in metal and in the Bose-condensate equal one to another:
    $E^{*}_{F} \sim 2.87 (h/2\pi)^{2}n_{cc}^{5/3}/m \sim  \Delta n_{cc}/2$  \cite{bib6,bib8}.
    An increase of \textit{x} up to 8 (or \textit{y} up to 1) transforms the oxide
    $\textrm{YBa}_{2}\textrm{Cu}_{3}\textrm{O}_{8-x}$ into the metal YBaCu. Only Cooper's pairs remain,
    while the most of electrons are in the Fermi degeneneracy state in result.

    Perhaps, such evolution of a system can take place in the noble gase (NG)
    condensation process \cite{bib6,bib7, bib8}. The concentration of virtually excited
    two-atom molecules with two paired electrons increases with compression,
    and the infinite cluster of chains of divalent particles is formed in the NG
    insulator (atoms in the ground state) above the percolation threshold:
    an irregular 3D conducting (superconducting) "cobweb" is created \cite{bib6}.
    Such a "cobweb" structure of the superconducting cluster "gossamer" \cite{bib9}
    can exist in HTS as well as it has been told above.
    Metallization of molecular condensates can be a useful model for study of the
    HTS materials. S and $\textrm{O}_{2}$ are metallized by now \cite{bib10,bib11}; their transition
    temperatures $T_{c}$ are respectively  $\sim 17\textrm{K}$ and 0.6K. Low temperatures can be
    due to strong local magnetic fields  of these paramagnetic atoms
    $(r_{O}=0.45$\!\!\AA; $r_{S}=0.81$\!\!\AA; $r_{Se}=0.92$\!\!\AA \, \cite{bib12}). \; Similar effect of the oxigen atoms
    can limit $T_{c}$ of the HTS oxides as well (Fig.1).

\paragraph*{\textbf{III}.}     Some physical property can be traced back to any of 4 sublattices of quantum
states. Sublattice (iv) is based on quantum ground state of $\textrm{O}\,(2p^{4})$ atoms, which have
two parallel spins. A presence of parallel spins determines paramagnetism of the
$\textrm{O}_{2}\, (2p^{2}-2p^{6})$ or $(\textrm{O}^{2+}-\textrm{O}^{2-})$ molecule.  Similar   molecule-like  state  exists  in  the
form of  dynamically  equilibrium  bonds  $(Me^{0} - \textrm{O}^{0} ) -- (Me^{2+} - \textrm{O}^{2-})$ in oxides. Below
the percolation threshold for superconductivity, there are almost $10^{22} \textrm{cm}^{-3}$ oxigen atoms
in the oxide; their paramagnetism is only partly compensated by metal atoms $(Me^{2+} - \textrm{O}^{2-}).$
Above the percolation threshold $y_{p},$ magnetic properties of the virtual sublattice of the
oxigen paramagnetic states turn to be compensated by increasing collective properties of
electrons of the cluster with increase of \textit{y} (Fig.1). Coexistence of superconductivity and
magnetism manifests a relation between physical properties and quantum states of atoms in oxides.
   Synthesis of materials free of atoms, which have real or at least virtual magnetic states,
   can be a perspective way of  the transition  temperature rise.  (For instance,
   $(M^{1})_{x}(M^{2})_{1-x}\textrm{B}_{6}$ with quantum states $2p^{0}-2p^{6}$ of B atoms in $\textrm{B}_{6}$
   molecule  or  $\textrm{B}^{1+}- \textrm{B}^{5-}$ bonds).
   Sulphides and selenides instead of oxides may be preferable for HTSC too.
 
 \begin{figure*}[tbp]
 \includegraphics[width=\linewidth]{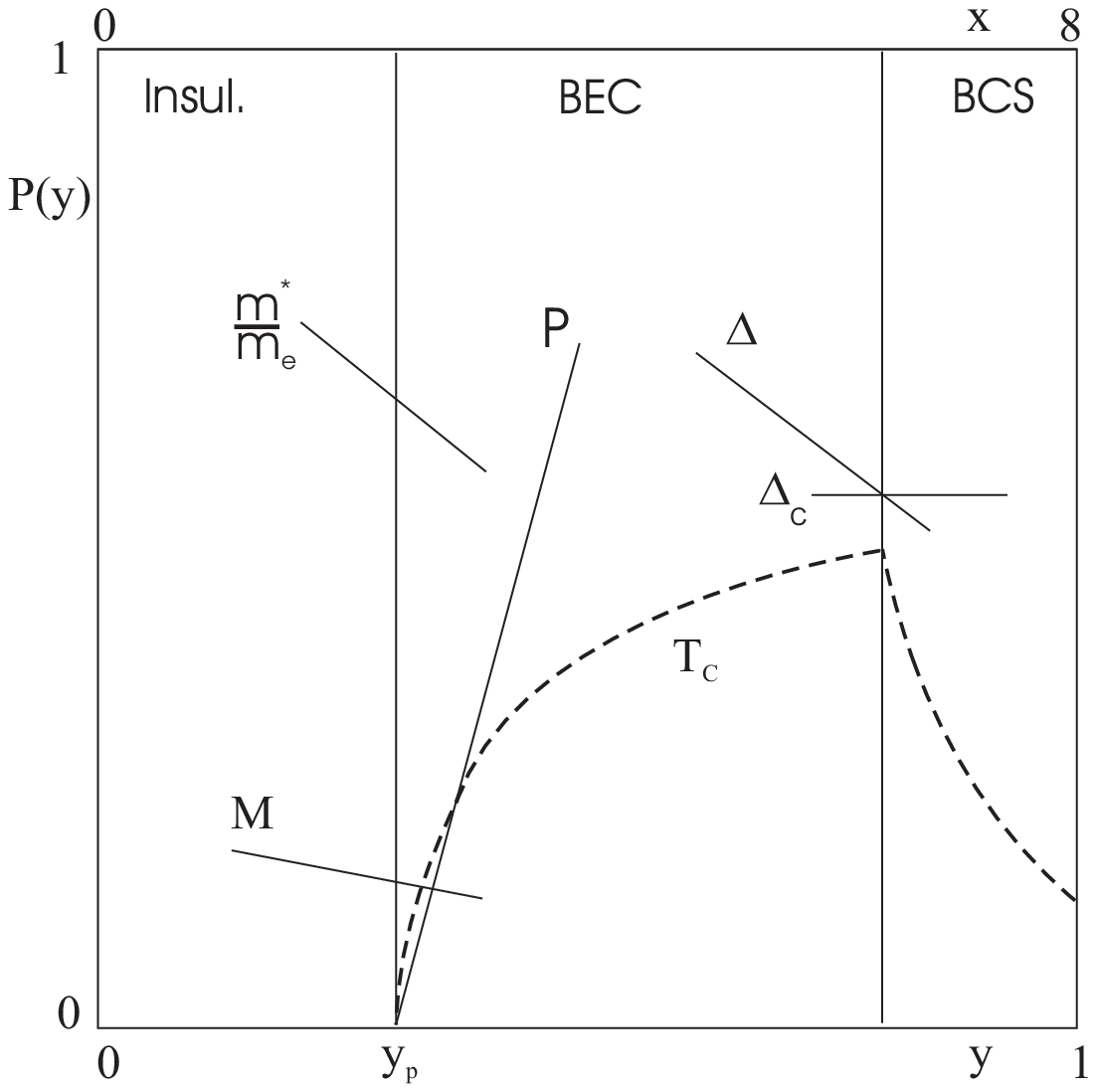}

 \caption{Schematic picture of some oxides properties\\      \textit{y} is the effective number of ground states per atom in the oxide;\\
 \textit{P(y)} is the effective number of ground states per atom of the metal in the
 superconducting cluster; \\
 M is the total magnetic moment of oxigen atoms $2p^{4}$ states in the oxide; \\
 $y_{p}$ is the percolation threshold for forming of the infinite superconducting cluster; \\
 $\Delta$ is the electron pair binding energy, $\Delta_{c}\sim 5.5 kT_{cB};$ \\
 \textit{x} is the parameter in the chemical formula $\textrm{YBa}_{2}\textrm{Cu}_{3}\textrm{O}_{8-x}.$}                                               \end{figure*}

\begin{thebibliography}{12}
 \bibitem{bib1} V.N.Bogomolov,       e-print http://xxx.lanl.gov/abs/cond-mat/0411574.
 \bibitem{bib2} G.Otto, E.Saur, H.Witzgall,        Journ. Low.Temp. Phys.   \textbf{1}, \textit{19} (1969).
 \bibitem{bib3} G.Bogner,           Electrotechn. Zs.   \textbf{89}, \textit{321}, (1968).
 \bibitem{bib4} B.Dam at al.,             Nature  \textbf{399}, \textit{439} (1999).
 \bibitem{bib5} G.Balestrino at al.,         Phys. Rev. Lett.  \textbf{89},  paper 156402 (2002).
 \bibitem{bib6} V.N.Bogomolov,  "Metallic xenon. Conductivity or
 superconductivity?"
   Preprint 1734, A.F.Ioffe PTI St-Peterburg (1999);  e-print
   http://xxx.lanl.gov/abs/cond-mat/9902353; /9912034.

 \bibitem{bib7} V.N.Bogomolov ,  Metallization of Molecular Condensates
 under Pressure as a Result of a Transition through the Percolation
 Threshold.
         Techn. Phys. Lett.,  21,  No. 11,  928 (1995).

 \bibitem{bib8} V.N.Bogomolov,     Techn.Phys.Lett.,  28, 211
 (2002).



 \bibitem{bib9}R.B.Laughlin,    e-print http://xxx.lanl.gov/abs/cond-mat/0209269.

 \bibitem{bib10} M.I.Eremets, E.A.Gregoryanz, V.V.Struzhkin, H.Mao,
                R.J.Hemley, N.Mulders,N.M.Zimmerman ,  Phys. Rev. Lett.  \textbf{85},  2797 (2000).
 \bibitem{bib11} K.Shimitsu, K.Suhara, M.Icumo, M.I.Eremets, K.Amaya ,    Nature, \textbf{393}, 767  (1998).
 \bibitem{bib12}  J.T.Waber, Don T.Cromer,       J.Chem. Phys.,   \textbf{42}, \textit{4116}  (1965).
 \end{thebibliography}
 \end{document}